\documentclass[%
 showpacs,
 reprint,%
 amssymb,amsmath,%
aip,jap,
floatfix]{revtex4-1}

\usepackage[latin1]{inputenc}

\newcommand{\muu}{\hbox{\textmu}}

\usepackage{lmodern}
\usepackage{textcomp}
\usepackage[T1]{fontenc}

\usepackage{graphicx}
\usepackage{dcolumn}
\usepackage{bm}
\usepackage{color}
\usepackage{amsfonts}
\usepackage{amsmath}
\newcommand{\notPart}[1]{}
\newcommand{\notPartData}[1]{}

\newcommand{\algaas}[0]{Al$_{0.3}$Ga$_{0.7}$As}

\begin{document}

\preprint{APS/123-QED}

\title{Investigations of Bragg reflectors in nanowire lasers}

\author{Guro K. Svendsen}
\author{Helge Weman}%
\author{Johannes Skaar}%
\altaffiliation[Also at ]{University Graduate Center, Kjeller, Norway.}

\affiliation{Department of Electronics and Telecommunications, Norwegian University of Science and Technology, NO-7491 Trondheim, Norway}
%


\date{\today}

\begin{abstract}
The reflectivity of various Bragg reflectors in connection to waveguide structures, including nanowires, has been investigated using modal reflection and transmission matrices. A semi-analytical model was applied yielding increased understanding of the diffraction effects present in such gratings. Planar waveguides and nanowire lasers are considered in particular. Two geometries are compared; Bragg reflectors within the waveguides are shown to have significant advantages compared to Bragg reflectors in the substrate, when diffraction effects are significant.
\end{abstract}

\pacs{
78.20.Ci, 
42.82.Et, 
42.81.Qb, 
81.07.Gf, 
42.55.Px, 
}
\maketitle
\section{Introduction}
Semiconductor nanowires including nanowire lasers are promising as building blocks for realization of nanoscale photonic devices \cite{mariano_rev, Duan2003241}. Various techniques such as molecular beam epitaxy (MBE) or metalorganic chemical vapour deposition (MOCVD) can be used to form nanowires with accurately controlled geometry and material composition, yielding a high level of flexibility \cite{Lauhon20041247}. To obtain an efficient laser resonator, the reflectivity at the end facet of the nanowire must be high. The refractive index contrast between the semiconductor nanowire and its surroundings is typically very large; the simplest nanowire-laser designs could thus use the cleaved end facets as reflectors. With such designs, the reflectivity of the guided lasing mode is quite moderate for single mode semiconductor nanowire lasers ($\sim25\%$ for GaAs-based nanowire, $\sim18\%$ for ZnO based) \cite{maslov, henneghien, svendsen:general}. Bragg gratings have been proposed to obtain a higher end facet reflectivity. Such gratings are fully compatible with most nanowire fabrication methods, e.g. MBE and MOCVD, and have already been realized experimentally\cite{Gudiksen2002617}. 
Chen \emph{et al.} have performed numerical analyses of nanowire Bragg structures. They show that a nanowire superlattice can be used to achieve near unity modal reflectivity at single mode operation \cite{chen_pbg}. Additionally, they have performed an optoelectronic analysis of nanowire lasers with distributed Bragg reflector mirrors \cite{chen_dbr}, showing a significant improvement in output power. Friedler \emph{et al.} \cite{friedler} have used coupled mode analysis to calculate the reflectivity of a dielectric Bragg grating within a GaAs nanowire. They conclude that the reflectivity of such a grating is rather poor in the single mode regime for a GaAs nanowire, and propose to rather use metallic mirrors. 

\begin{figure}%
\includegraphics[width=\columnwidth]{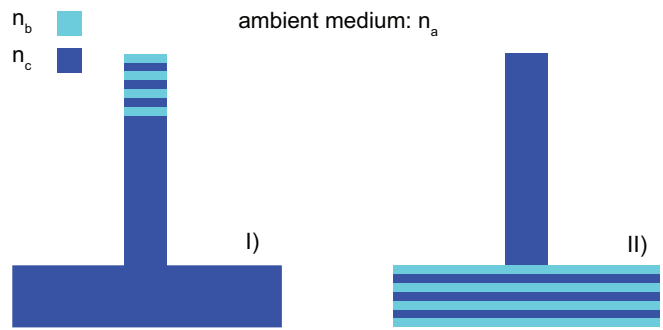}%
\caption{Two geometries for implementation of Bragg mirror at the end facets of a nanowire laser.}%
\label{fig:transfer_geom}%
\end{figure}

When the lateral scale of a waveguide is of the order of the wavelength of the guided light, diffraction effects become significant. In this work we perform a detailed analysis of Bragg grating reflectors in connection to such diffractive waveguides, to investigate in which regimes a Bragg grating is efficient. A semi-analytical model will be used; as compared to finite element methods this helps in explaining more of the mechanisms that influence the reflectivity. A substrate grating (Fig.~\ref{fig:transfer_geom} II) is found to have surprisingly low reflectivity compared to a grating within the waveguide (Fig.~\ref{fig:transfer_geom} I). Furthermore we see that even for extremely small waveguides, where only a small fraction of the field is within the waveguide, a near unity reflectivity may be obtained by having enough periods in the Bragg grating.  

The reflection and transmission properties of an interface are fully described by reflection and transmission matrices. These matrices describe the amount of mode $i$ that is reflected or transmitted into mode $j$.  In our previous work a formalism was developed to calculate the transmission and reflection matrices for end facets of waveguides \cite{svendsen:general}. The method is particularly useful for highly diffractive waveguides and nanowire laser applications. In this paper we extend the formalism to examine the effect of Bragg reflectors, and consider two different geometries; (I) Bragg reflector in the substrate, (II) Bragg reflector at the top of the nanowire (within the waveguide).	The two geometries are sketched in Fig.~\ref{fig:transfer_geom}. A heterostructure based on GaAs and \algaas{} is used throughout as an example.

The outline of this article is as follows: A presentation of the multimode transfer matrix formalism is presented in Sec.~\ref{sec:transfer}. The calculation model for the reflection and transmission matrices is presented in Sec.~\ref{sec:reflection_transmission}. A brief summary of previous work describing the reflection at the end facet of a waveguide is given, as well as the necessary generalizations to describe a Bragg grating within a waveguide or in the substrate. Sec.~\ref{sec:bragg_design} contains a discussion concerning the design of the Bragg gratings. Numerical results for a planar waveguide structure are given in Sec.~\ref{sec:results_planar}, and some results concerning nanowires, with 2D confinement are given in Sec.~\ref{sec:results_nanowire}.

\section{Transfer matrix formalism}
\label{sec:transfer}
The theory of transfer and scattering matrices can be found in standard textbooks \cite{Saleh_teich}. We will here briefly review the concepts, to introduce our choice of notation. Consider a stack of layers, with layer boundaries perpendicular to the propagation axis, $z$. Each layer is homogeneous w.r.t. $z$. The field in each layer can be described using its modes.  Throughout this article we define modes as being pairs of electric and magnetic fields that are eigenfunctions of the electromagnetic propagation operator along the $z$-axis. For an infinitely long waveguide that is homogeneous along the $z$-axis, the modes will correspond to the eigenmodes of the whole structure. However, for waveguides of finite length or with inhomogeneities the modes are merely local modes, not to be confused with the supermodes of the overall structure. 

Let the forward propagating mode $n$ in layer $i$ have amplitude $a^i_n$, and the backward propagating mode have amplitude $b^i_n$. The vectors $a^i$ and $b^i$ contain the amplitudes of all forward propagating modes and backward propagating modes, respectively.
Let $r^{ji}$ and $t^{ji}$ be matrices describing the modal reflection and transmission respectively, for light incident from layer $i$ towards layer $j$; similarly $r^{ij}$ and $t^{ij}$ describe the reflection/transmission from the opposite side. Using these matrices, we can relate the field in layer $i$ to the field in layer $j$:
\begin{subequations}
\label{eq:neighbour_rel}
\begin{align}
b^i_k=\sum_{l}{r^{ji}_{kl}a^i_l}+\sum_{l'}{t^{ij}_{kl'}b^j_{l'}}\\
a^j_k=\sum_{l}{t^{ji}_{kl}a^i_{l}}+\sum_{l'}{r^{ij}_{kl'}b^j_{l'}}.
\end{align}
\end{subequations}
The matrix $r^{ji}$ has elements $r_{kl}^{ji}$, i.e. $r^{ji}=\left[r_{kl}^{ji}\right]$, similarly for  
$t^{ji},r^{ij}$ and $t^{ij}$.
Eq.~\eqref{eq:neighbour_rel} can be rewritten in matrix form as
\begin{eqnarray}
\left[ \begin{array}{c}
b^i\\
a^j\end{array} \right]=S^{ji}
\left[ \begin{array}{c}
a^i\\
b^j\end{array} \right].
\label{eq:scattering matrix_1}
\end{eqnarray}
Here $S^{ji}$ is the scattering matrix:
\begin{eqnarray}
S^{ji}=\left[ \begin{array}{c c}
r^{ji}& t^{ij}\\
t^{ji}& r^{ij}\end{array} \right].
\label{eq:scattering_matrix}
\end{eqnarray}

When considering a sequence of layers, it is convenient to reformulate \eqref{eq:scattering matrix_1} so that the field in layer $j$ can be explicitly expressed using the field in layer $i$, i.e.,
\begin{eqnarray}
\left[ \begin{array}{c}
a^j\\
b^j\end{array} \right]=M^{ji}
\left[ \begin{array}{c}
a^i\\
b^i\end{array} \right].
\label{eq:transfer_def}
\end{eqnarray}
The matrix $M^{ji}$ is known as the transfer matrix; a general transfer matrix is illustrated in Fig.~\ref{eq:transfer_matrix}. In light of \eqref{eq:neighbour_rel} it can be expressed in terms of the scattering matrix $S^{ji}$:
\begin{eqnarray}
M^{ji}=\left[ \begin{array}{c c }
t^{ji}-r^{ij}\left(t^{ij}\right)^+r^{ji},&\,r^{ij}\left(t^{ij}\right)^+\\
-\left(t^{ij}\right)^+r^{ji},&\,\left(t^{ij}\right)^+\end{array} \right].
\label{eq:transfer_matrix}
\end{eqnarray}
\begin{figure}%
\includegraphics[width=\columnwidth]{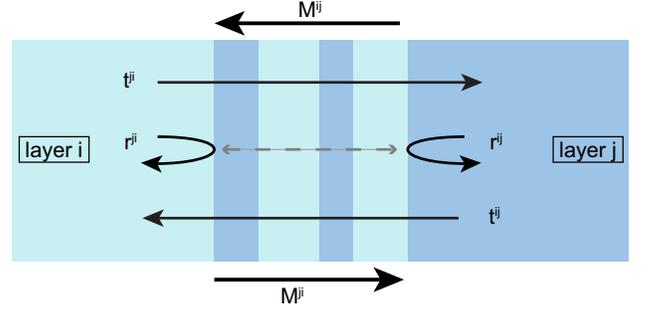}%
\caption{The transfer matrix $M^{ji}$ relates the field in layer $i$ to the field in layer $j$.}%
\end{figure}
Here, the superscript $(+)$ denotes the matrix inverse or More-Penrose pseudo inverse depending on whether the matrix is quadratic or rectangular. 
The presence of evanescent modes in a layer could cause ill-conditioned transfer matrices. As the transmission coefficients $t^{ij}_{kl}$ involving evanescent modes may be extremely small, matrix inversion of the transmission matrix may cause numerical instabilities. To avoid such problems it is preferable to use recursive relations derived from transfer matrices rather than direct matrix multiplication. We consider a stack of three layers, 1,2 and 3; the individual transfer matrices are multiplied to find the total reflection and transmission properties. Recall that $r^{ji}$( $t^{ji}$) denote the reflection (transmission) from layer $i$ to layer $j$. The combined reflection and transmission coefficients for the system of layers are given by:

\begin{subequations}
\label{eq:recursive}
\begin{align}
&r^{31}=r^{21}+t^{12}\left(I-r^{32}r^{12}\right)^+r^{32}t^{21}\\
&t^{31}=t^{32}\left(I+r^{12}\left(I-r^{32}r^{12}\right)^+r^{32}\right)t^{21}\\
&r^{13}=r^{23}+t^{32}r^{12}\left(I-r^{32}r^{12}\right)^+t^{23}\\
&t^{13}=t^{12}\left(I-r^{32}r^{12}\right)^+t^{23}.
\end{align}
\end{subequations}

Propagation in the $z$-direction within one layer can be described in the same manner. Mode $k$ propagates according to 
\begin{subequations}
\label{eq:propagation_rel}
\begin{align}
b^i_k=\textrm{e}^{i\beta_{k}d}b^j_k\\
a^j_k=\textrm{e}^{i\beta_{k}d}a^i_{k},
\end{align}
\end{subequations}
where $\beta_k$ is the modal propagation constant in $z$-direction, and $d$ is the propagation distance.

\section{Finding the reflection and transmission matrices}
\label{sec:reflection_transmission}
\begin{figure}%
\includegraphics[width=\columnwidth]{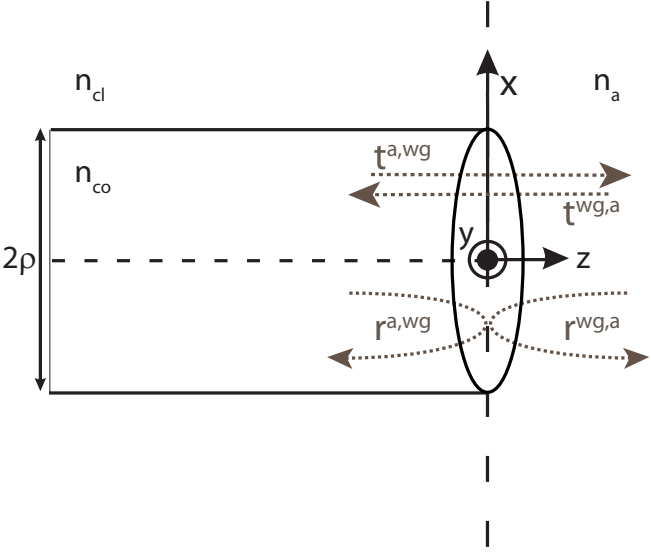}%
\caption{Circular waveguide of diameter $2\rho$ oriented along the $z$-axis with an end facet at $z=0$. The waveguide core and cladding have refractive indices $n_\text{co}$ and $n_\text{cl}$, respectively. For $z$>0 there is an ambient medium, which is assumed to be homogeneous with refractive index $n_\text{a}$. The aim is to determine the reflection and transmission matrices for the end facet, as illustrated in the figure.} %
\label{fig:geom_circ_wg_a}%
\end{figure}

The problem of finding the reflection and transmission matrices at the end facet of a waveguide terminated in a homogeneous medium has been addressed by us previously\cite{svendsen:general}, here we briefly sum up the main results.
The geometry of the problem is shown in Fig.~\ref{fig:geom_circ_wg_a}, here exemplified using a circular waveguide. 
We describe the field at both sides of the boundary, $z=0$, using a set of modes. The modes in the half-space z$>$0, constitute a continuous set of radiation modes, whereas for z$<$0 the modal spectrum consists of a discrete set of bound modes and a continuous set of radiation modes. The modal spectrum is discretized using periodic boundary conditions at each side, the width of the computational cell in both $x$ and $y$ direction is $2L$. The electric field of mode $m$ in the ambient half space can be written as
\begin{align}
&\mathbf{\mathcal{E}}_{m}(x,y)= \mathbf{\hat{\mathcal{E}}}_{m}\exp(ik_xx+ik_yy).\label{modeexpE}
\end{align}
The magnetic field, $\mathbf{\mathcal{H}}_{m}$, is described in the same way.
The label $m$ is a collection of the modal indices, $m=(p,q,\textrm{pol})$. The polarization, pol, is TE or TM, and the real transverse wavevectors are $k_x=p\frac{\pi}{L}$, $k_y=q\frac{\pi}{L}$, where $p$ and $q$ are integers. The modal propagation constant $k_z$ is given by $k_z^2=n_{\text{a}}^2\omega^2/c^2-k_x^2-k_y^2$, where $n_{\text{a}}$ is the refractive index of the half-space $z>0$, and $c$ is the vacuum light velocity.  
The constant vectors can be expressed \footnote{In the limit $k_x=k_y\to0$, the corresponding limit of the expressions in \eqref{eq:constvectors} must be used}
\begin{subequations}\label{eq:constvectors}
\begin{align}
&\mathbf{\hat {\mathcal{E}}}_{m(\text{TE})}=A\sqrt{\omega\mu}(-k_y,k_x,0),\label{eTE}\\
&\mathbf{\hat {\mathcal{E}}}_{m(\text{TM})}=\frac{A}{\sqrt{\omega\varepsilon_{\text{a}}}}\left(k_xk_z,k_yk_z,-\left(k_x^2+k_y^2\right)\right),\label{eTM}\\
&\mathbf{\hat {\mathcal{H}}}_{m(\text{TE})}=\frac{A}{\sqrt{\omega\mu}}(-k_xk_z,-k_yk_z,k_x^2+k_y^2),\label{hTE} \\
&\mathbf{\hat {\mathcal{H}}}_{m(\text{TM})}=A\sqrt{\omega\varepsilon_{\text{a}}}(-k_y,k_x,0)\label{hTM},
\end{align}
\end{subequations}
where $A=1\left/{\sqrt{(k_x^2+k_y^2)|k_z|2L^2}}\right.$. We assume that the medium is nonmagnetic, and the permittivity of the medium is $\varepsilon_{\text{a}}$.
The modal fields of the waveguide are denoted ${\mathbf e_i=\mathbf{e}_i(x,y)}$ and $\mathbf h_i=\mathbf h_i(x,y)$, $i=1,2,\ldots$.
We now use the continuity of the transverse electric and magnetic fields. Assuming the incoming mode $\{\mathbf{e}_i,\mathbf h_i\}$, we can write
\begin{subequations}
\label{eq:bound_cond}
\begin{align}
&\mathbf e_i^{(t)}+\sum_j r_{ji}^\text{a,wg}\mathbf e_j^{(t)}=\sum_{m} t_{mi}^\text{a,wg}\mathbf{\mathcal{E}}_{m}^{(t)}\label{contE}\\
&\mathbf h_i^{(t)}-\sum_j r_{ji}^\text{a,wg}\mathbf h_j^{(t)}=\sum_{m} t_{mi}^\text{a,wg}\mathbf{\mathcal{H}}_{m}^{(t)}\label{contH},
\end{align}
\end{subequations}
valid for all $x$ and $y$. Here $r_{ji}^\text{a,wg}$ is the reflection coefficient from mode $i$ to mode $j$, and $t_{mi}^\text{a,wg}$ is the transmission coefficient from mode $i$ ($z<0$) to mode $m$ ($z>0$). The superscript $(t)$ stands for the transverse component ($x$ and $y$ components) of the vector. 
Eqs. \eqref{contE} and \eqref{contH} can be combined as follows. Take the vector product between \eqref{contE} and $\mathbf{\mathcal{H}}_{m'}^{(t)*}(x,y)$, and integrate over the unit cell. Similarly, take the vector product between $\mathbf{\mathcal{E}}_{m'}^{(t)*}(x,y)$ and \eqref{contH}, and integrate over the unit cell. Combining the resulting equations yield
\begin{subequations}
\label{eq:r_t_front}
\begin{align}
&r^\text{a,wg}=[r_{ji}]^\text{a,wg}=[\Phi_{im}-\Psi_{im}][\Phi_{im}+\Psi_{im}]^{+},\label{eq:reflection}\\
&t^\text{a,wg}=[t_{mi}]^\text{a,wg}=\frac{1}{2}\left([\Phi_{mi}+\Psi_{mi}]-[\Phi_{mi}-\Psi_{mi}]r^\text{a,wg}\right)\label{eq:transmission}
. 
\end{align}
\end{subequations}
Here we have defined the inner products
\begin{subequations}\label{eq:innerprod}
\begin{align}
&\Psi_{mi}=\frac{\kappa(m)^*}{|\kappa(m)|}\frac{1}{2}\int\limits_{\text{cell}}\mathbf e_i^{(t)}\times\mathbf{\mathcal{H}}_{m}^{(t)*}(x,y)\cdot\mathbf{\hat z}\mathrm{d}A, \\
&\Phi_{mi}^*=\frac{\kappa(m)^*}{|\kappa(m)|}\frac{1}{2}\int\limits_{\text{cell}}\mathbf{\mathcal{E}}_{m}^{(t)}(x,y)\times\mathbf h_i^{(t)*}\cdot\mathbf{\hat z}\mathrm{d}A.
\end{align}
\end{subequations}
The unit vector in the $z$-direction is $\mathbf{\hat z}$, and
\begin{equation}\kappa(m)=\left\{ \begin{smallmatrix} 
k_z^*,\:\text{pol}=\text{TE}\\ 
k_z,\:\text{pol}=\text{TM}. \end{smallmatrix} \right.\end{equation}

It is straightforward to extend the formalism to describe the reflection and transmission for light incident onto the facet from the ambient medium. We assume the incoming wave $\{\mathbf{\mathcal{E}}_i,\mathbf{\mathcal{H}}_i\}$, and consider the boundary conditions, similarly to  \eqref{eq:bound_cond}. The resulting expressions for the reflection and transmission are
\begin{subequations}\label{eq:r_t_back}
\begin{align}
&r^\text{wg,a}=[r_{ji}]^\text{wg,a}=\left([\bar{\Psi}_{mi}+\bar{\Phi}_{mi}]^{*}\right)^+[\bar{\Psi}_{mi}-\bar{\Phi}_{mi}]^{*}\label{eq:reflection_back}\\
&t^\text{wg,a}=[t_{mi}]^\text{wg,a}=\frac{1}{2}\left([\bar{\Phi}_{mi}+\bar{\Psi}_{mi}]^{*}-[\bar{\Psi}_{mi}-\bar{\Phi}_{mi}]^{*}r^\text{wg,a}\right). 
\end{align}
\end{subequations}
Here,
\begin{subequations}\label{eq:innerprod_back}
\begin{align}
&\bar{\Psi}_{im}=\frac{\beta_m}{|\beta_m|}\frac{1}{2}\int\limits_{\text{cell}}\mathbf e_m^{(t)}\times\mathbf{\mathcal{H}}_{i}^{(t)*}(x,y)\cdot\mathbf{\hat z}\mathrm{d}A, \\
&\bar{\Phi}_{im}^*=\frac{\beta_m}{|\beta_m|}\frac{1}{2}\int\limits_{\text{cell}}\mathbf{\mathcal{E}}_{i}^{(t)}(x,y)\times\mathbf h_m^{(t)*}\cdot\mathbf{\hat z}\mathrm{d}A.
\end{align}
\end{subequations}
We have here assumed that the waveguide modes are orthogonal, and fulfill 
\begin{equation}
\frac{1}{2}\int_{\text{cell}} \mathbf{e}^j\times\mathbf{h}^{j'*}\cdot\mathbf{\hat z} dA=\frac{\beta^{j*}}{|\beta^j|}\delta_{jj'}.
\label{orthonormalwaveguide}
\end{equation}
This orthonormality relation can always be fulfilled for the modes of nonabsorbing waveguides \cite{snyder}. 
Note that the transmission matrices may also be found directly from the inner products;
\begin{subequations}\label{eq:transmission_dir}
\begin{align}
&t^\text{wg,a}=2\left({{[\bar{\Psi}_{im}]}^{*}}^++{[\bar{\Phi}_{im}]^{*}}^+\right)^+,\\ 
&t^\text{a,wg}=2\left([\Phi_{mi}]^++{[\Psi_{mi}]}^+\right)^+.
\end{align}
\end{subequations}

A mode is said to be real if it has a real-valued propagation constant and the transverse electric and magnetic field of the mode can be written real for all values of $x$ and $y$. For modes in nonabsorbing waveguides the transverse fields can always be written real \cite{snyder}. Modes with a real-valued propagation constant will therefore be real modes. For coupling between  real modes $i$ and $m$ with real propagation constants, we have $\beta_i/|\beta_i|=\kappa(m)^*/|\kappa(m)|$.  In this case $\Psi_{mi}$ and $\bar{\Psi}_{im}$ are both real, and we have $\Psi_{mi}=\bar{\Psi}_{mi}$, and $\Phi_{mi}=\bar{\Phi}_{mi}$. We then see directly that $t^\text{a,wg}=t^\text{wg,a}$, exactly as predicted by the reciprocity theorem.

It is a necessary condition when solving for the reflection matrices to have a well defined system of equations. A minimum requirement is to use the same number of orthogonal modes at both sides of the boundary. This is however not an ideal solution, as the sampling in the spatial frequency domain is quite different at the two sides of the boundary; thus a large number of modes would be necessary for an accurate description of the interface. We have rather chosen to use a higher number of modes on the ambient side; this enables a good description of the forward reflection $r^\text{a,wg}$. However, for the backward reflection $r^\text{wg,a}$, we cannot directly find the reflection coefficients for all modes of the ambient medium. The procedure is as follows: First we find the reflection matrix for the ambient modes with the lowest spatial frequencies. These modes must be sufficiently well described by a superposition of waveguide modes. More precisely they obey $\left(k_x^2+k_y^2\right)\leq \left(\left(n_\text{co}\omega/c\right)^2-\beta_\text{lim}^2\right)$, where $\beta_\text{lim}$ is the possibly imaginary propagation constant of the highest order waveguide mode. For higher spatial frequencies, we approximate the reflection coefficients using the scalar Fresnel equations for  reflection  at an interface with index contrast $n_\text{a}/n_\text{cl}$. This approximation shows very good agreement provided the number of modes on the waveguide side is not too small.

For Bragg gratings within the waveguide one also needs to describe the reflection and transmission properties when there is a waveguide at both sides of the interface. One possibility for performing this calculation would be to repeat the procedure described previously, using waveguide modes at each side of the interface. As we have waveguide  modes at both sides of the boundary the inner products similar to \eqref{eq:innerprod} and \eqref{eq:innerprod_back} could no longer be formulated as Fourier transforms. To avoid this problem, we formulate the reflection and transmission matrices for transitions between waveguides in terms of the previously acquired relations for a transition from a waveguide to an ambient \eqref{eq:r_t_front}. We start by formulating the boundary condition as in \eqref{eq:bound_cond}. The fields at both sides are then expressed in terms of the inner products \eqref{eq:innerprod} between each waveguide and a dummy ambient layer. Using this procedure we obtain expressions for the reflection and transmission between waveguides formulated in terms of their reflection and transmission matrices towards a dummy ambient medium. Note that there are no assumptions made here, and the accuracy is given from the accuracy of the reflection and transmission matrices from the waveguide to the dummy ambient medium. The details concerning this calculation are given in Appendix \ref{sec:bragg_within}. For a transition from waveguide $b$ towards waveguide $c$ the result is

\begin{subequations}
\label{eq:wgc_wgb_brief}
\begin{align}
&t^{cb}=G^{cb}\left(I-{r^{ab}}r^{cb}\right)\\
&r^{cb}=\left(G^{cb}-r^{ac}G^{cb}r^{ab}\right)^+\left(G^{cb}r^{ab}-r^{ac}G^{cb}\right)
,\end{align}
\end{subequations}
where the superscript $a$ denotes the dummy ambient media, and
\begin{subequations}
\begin{align}
G^{cb}=\left(I-r^{ac}r^{ac}\right)\left(t^{ac}\right)^+t^{ab}\left(I-r^{ab}r^{ab}\right)^+.\nonumber\\
\end{align}
\label{eq:G_bc_brief}
\end{subequations}
The opposite transition is described by interchanging indices $b$ and $c$.

For a thin diffractive waveguide, the imposed boundary conditions cause artificial reflections from the boundary. To deal with these artificial reflections, we introduce some loss into the system, i.e., $\varepsilon\to\varepsilon+i\gamma\varepsilon_0$, at both sides of the boundary. The loss parameter $\gamma$ should be small enough not to alter the reflection properties of the boundary significantly \cite{svendsen:general}. 
Note that this loss is merely artificial, and it will only be included when necessary. To describe the Bragg grating, we must treat the scattering at interfaces as well as propagation in homogeneous layers. The loss is not included in the propagation description, as this would lead to an underestimate of the reflection as compared to the physical situation. Since the loss is included in the description of the interfaces, but not in the propagation description, it represents a deviation from a physical structure and some error is to be expected in the final result. Decreasing $\gamma$ will decrease this error. We previously assumed that the waveguide modes fulfilled the orthonormality relation \eqref{orthonormalwaveguide}. This is however only generally true for nonabsorbing waveguides. For slightly absorbing waveguides we may assume that the deviation from \eqref{orthonormalwaveguide} is small\cite{snyder}. The orhonormality relation can even be exactly fulfilled for the planar step index waveguides considered in this paper when $\varepsilon\to\varepsilon+i\gamma\varepsilon_0$.

\section{Designing the Bragg grating}
\label{sec:bragg_design}
In this section we consider the design of the Bragg grating structure in connection to a waveguide. Let a waveguide be terminated by some grating consisting of layers with refractive indices $n_\text{b}$ and $n_\text{c}$, and thicknesses $d_\text{b}$ and $d_\text{c}$ respectively. The waveguide itself has refractive index $n_\text{c}$. The structure is designed to be a quarter wave stack for the fundamental mode of the waveguide, i.e. the mode with propagation constant $\beta^{(1)}_\text{c}$ in $z$-direction. The thicknesses of the quarter wave layers are given by
\begin{subequations}
\begin{align}
d_\text{c}=\frac{2\pi}{4\beta^{(1)}_\text{c}}\\
d_\text{b}=\frac{2\pi}{4\beta^{(1)}_\text{b}}
\end{align}
\end{subequations}

The response of the grating is highly dependent on which material that constitutes the terminating layer. To illustrate this we consider a planar structure with the grating within the waveguide (Fig.~\ref{fig:transfer_geom} I).  The heterostructure is based on GaAs and Al$_{0.3}$Ga$_{0.7}$As and the ambient medium is vacuum. The waveguide where the lasing is to occur consists of GaAs, i.e. $n_\text{c}=n$(GaAs), $n_\text{b}=n$(AlGaAs). The lasing wavelength for GaAs in the Zinc blende (ZB) crystal phase is 870 nm at room temperature \cite{Vurgaftmanband_param}. At this wavelength, the refractive indices of  GaAs and Al$_{0.3}$Ga$_{0.7}$As are\cite{jenkins:1848} $n\left(\text{GaAs}\right)=$3.6 and $n\left(\text{AlGaAs}\right)=$3.4. The structure is surrounded by air.  The total reflection matrix is found using the recursive relations \eqref{eq:recursive}. To this end we need the propagation matrices, the reflection and transmission matrices describing the interfaces between the waveguide layers, and the reflection matrix for the transition from the terminating grating layer towards the surrounding ambient.  

\begin{figure}%
\includegraphics[width=\columnwidth]{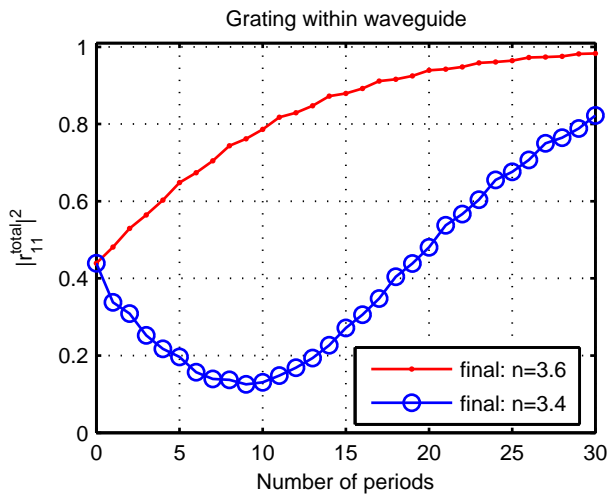}%
\caption{Reflection coefficient $|r_{11}^{\text{total}}|^2$ of a GaAs/\algaas{} Bragg grating in a planar waveguide, with heterostructure consisting of media with $n=3.4$ and $n=3.6$. The structure is surrounded by vacuum. The figure shows the reflection coefficient when either the low index material or the high index material is the final layer of the structure. Here, $a(\omega/c)=1$.}
\label{fig:design}%
\end{figure}

We calculate the total reflection when the terminating layer consists of either the low index or the high index material. Let the frequency of the light be $\omega$ and the width of the waveguide be $2a$. We include all modes with $\beta^2>\beta_\text{lim}^2$, where $\beta_\text{lim}$ is the cut-off limit. In this example, $a(\omega/c)=1$, $\beta_\text{lim}=10i(\omega/c)$, $L=100 a$, and $\gamma_i=0.1$. Fig.~\ref{fig:design} shows the reflection coefficient for the fundamental TE even mode, as a function of the number of periods, for gratings terminated by either GaAs or \algaas{}. Note that the behavior is fundamentally different depending on whether the grating is terminated by the high index material or by the low index material. When the material with the lowest refractive index terminates the grating, the reflection is reduced rather than increased for the first layers. 

This can be explained as follows. For a quarter wave stack all multiple reflections interfere constructively, as there is an additional phase shift of $\pi$ at every second interface when the refractive index goes from low to high. If the interface towards the ambient layer breaks this periodicity, the portion of the field reflected at this last interface will interfere destructively with the rest of the field. Unlike in conventional Bragg gratings, this last reflection may be crucial, as there is such a large index contrast between the grating and the surrounding air. If the grating consists of several periods, most of the field is reflected before it reaches the last interface; the effect of this additional phase shift is therefore gradually reduced. Fig.~\ref{fig:design} clearly shows that to enable an efficient grating the terminating layer should be made from the high index material. An alternative solution if one needs to terminate the grating using the lowest index material, is to grow the terminating layer with twice the thickness to compensate for the phase shift. A waveguide grating terminated by the highest-index material followed by an ambient of the same material would yield a similar effect, as the effective refractive index of the fundamental mode in the waveguide is lower than the refractive index of the bulk material. This may be part of the reason why Friedler \emph{et al.}~\cite{friedler} obtain so low reflectivity for the thinnest waveguides. As the thickness of the waveguide increases, the index contrast and thus the reflectivity at the last interface will decrease, and this effect would diminish.

The phase shift at the first interface has a similar effect. If the first interface breaks the periodicity, the contribution from this first reflection will be out of phase with the remaining contributions. This situation may e.g. occur when a GaAs waveguide is terminated by an AlGaAs/GaAs Bragg grating in the substrate (Fig.~\ref{fig:transfer_geom} II). As the waveguide thickness decreases below a certain limit, the effective refractive index of the fundamental mode in the waveguide will decrease below that of the first layer of the substrate (consisting of AlGaAs). This will cause the reflection at the first interface to interfere destructively with the remaining backscattered field, leading to reduced reflectivity. One possible solution to compensate for this phase shift, is to adjust the thickness of the first layer accordingly. 


\section{Planar waveguide structure}
\label{sec:results_planar}
We now look more closely into some numerical examples for a planar waveguide with a Bragg grating. Two situations are considered; the grating is either within the waveguide (Fig.~\ref{fig:transfer_geom} I) or in the substrate below the waveguide  (Fig.~\ref{fig:transfer_geom} II)). We also briefly consider an intermediate geometry. The planar waveguide with 1D confinement is less computationally demanding compared to the 2D case; in addition both bound and unbound modes can be found analytically \cite{svendsen:general}. The planar case is therefore well suited to test qualitative relations and convergence criteria. In a planar waveguide there is no coupling between modes of different parity (odd/even) or between modes of orthogonal polarization (TE/TM). The discussion is therefore limited to even TE-polarized modes. 

First, let the Bragg grating be within the waveguide, as shown in Fig.~\ref{fig:transfer_geom} I). Such structures can be realized by growing the Bragg grating at the end of the nanowire growth, by alternating the source materials during the epitaxial growth. In this example, the main part of the waveguide consists of GaAs, and alternating layers of AlGaAs and GaAs are grown at the top of the waveguide. The uppermost layer consists of GaAs, and the structure is surrounded by vacuum. We have performed calculations for up to 100 periods, to see the behavior in the limit of several periods. Note however that this is a very high number, which is not easily achieved with today's technology.  

\begin{figure}%
 \includegraphics[width=\columnwidth]{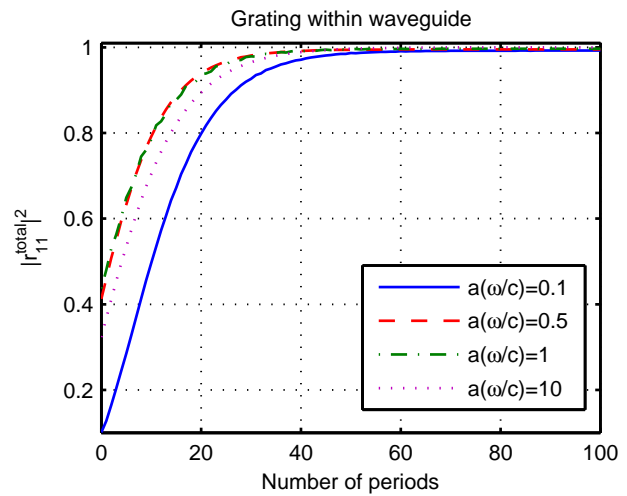}%
\caption{The reflection coefficient $\left|r_{11}^\text{total}\right|^2$ as a function of the number of periods in the GaAs/Al$_{0.3}$Ga$_{0.7}$As Bragg grating, for four waveguides of various width, $a$.}.
\label{fig:bragg_top_01}%
\end{figure}

We consider four normalized waveguide widths; $a(\omega/c)=0.1$, $a(\omega/c)=0.5$, $a(\omega/c)=1$, and $a(\omega/c)=10$. For reference, the single mode regime for even TE modes in this GaAs waveguide extends up to $a=0.91(\omega/c)$. 
The computational cell half-width $L$ is $100/(\omega/c)$,  and the loss parameter is $\gamma=0.1$. The cut-off limits $\beta_\text{lim}$ were chosen by considering the convergence of the reflection in each case. In the order of increasing waveguide width we used $\beta_\text{lim}=20(\omega/c)i$, $\beta_\text{lim}=15(\omega/c)i$, $\beta_\text{lim}=10(\omega/c)i$, and $\beta_\text{lim}=5(\omega/c)i$. 
The reflection coefficient $\left|r_{11}^\text{total}\right|^2$, i.e. the amount of the fundamental mode reflected back into itself, is shown in Fig.~\ref{fig:bragg_top_01} as a function of the number of periods in the Bragg grating.

It is seen that the end facet reflectivity of all waveguides converge towards a value very close to 1. Even as the normalized waveguide width decreases below $a\omega/c=0.5$, one can still obtain high reflectivity by increasing the number of periods in the grating. As will be seen later this is contrary to what is observed for the case with the grating in the substrate. 

Before we proceed it is instructive to review somewhat how the modal fields are influenced by diffraction. Firstly, as the width of the waveguide decreases, a decreasing proportion of the modal field will be confined within the core of the waveguide. As a consequence, the effective refractive index of the fundamental mode decreases towards the limit where it is close to the refractive index of the cladding material. Secondly, as the modes are confined to smaller areas in space, a corresponding spreading of the spatial frequencies of the mode must follow. This will e.g. imply that the waveguide modes will couple more strongly to each other upon reflection, and that there will be a larger angular spread of the beam upon transmission towards an ambient medium.
 
\begin{figure}%
\includegraphics[width=\columnwidth]{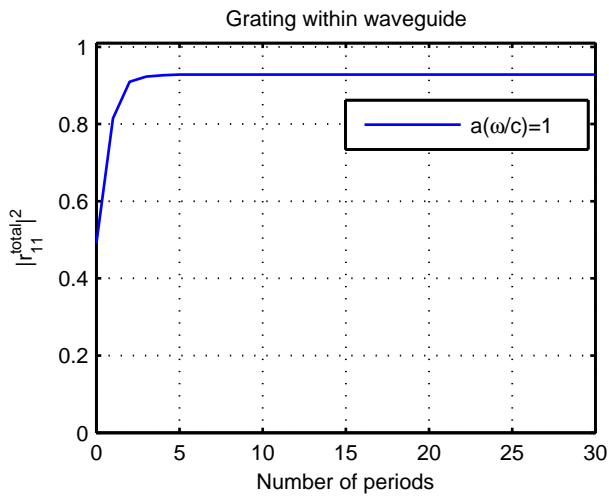}%
\caption{The reflectivity $\left|r_{11}^\text{total}\right|^2$ as a function of the number of periods in the Bragg grating for a waveguide of normalized width $a/(\omega/c)=1$. The Bragg grating consists of two materials with refractive indices $n=4$ and $n=2$.}%
\label{fig:bragg_top_a1_large_contrast}%
\end{figure}

A grating within the waveguide with a relatively low refractive index contrast will roughly preserve the same set of modes along the grating. Except for the modes that are very close to their cut-off, each mode will thus experience a jump in the effective refractive index in a manner quite similar to what is seen for conventional Bragg gratings. Only a small amount of energy will therefore be transferred from e.g. the fundamental mode to the higher order modes. If the index contrast of the grating is larger, each transition in the grating represents a more significant perturbation to the modal field, and there will be a larger amount of cross coupling between modes. Some of the energy from the fundamental mode may thus couple into other modes. Fig.~\ref{fig:bragg_top_a1_large_contrast} displays the reflectivity of a Bragg grating consisting of two materials with higher index contrast. Here GaAs has been replaced by a material with refractive index 4, and AlGaAs has been replaced by a medium of refractive index 2. The simulation parameters are the same as for the corresponding GaAs/\algaas{} structure. Note that the reflection coefficient of the fundamental mode now converge towards a value less than unity. As can be seen by comparing Fig.~\ref{fig:bragg_top_01} and Fig.~\ref{fig:bragg_top_a1_large_contrast}, there is a trade-off here in terms of the index contrast. Higher index contrast enables quite high reflection using fewer periods. On the other hand the maximum obtainable reflection is larger for the lower index contrast system.

\begin{figure}%
\includegraphics[width=\columnwidth]{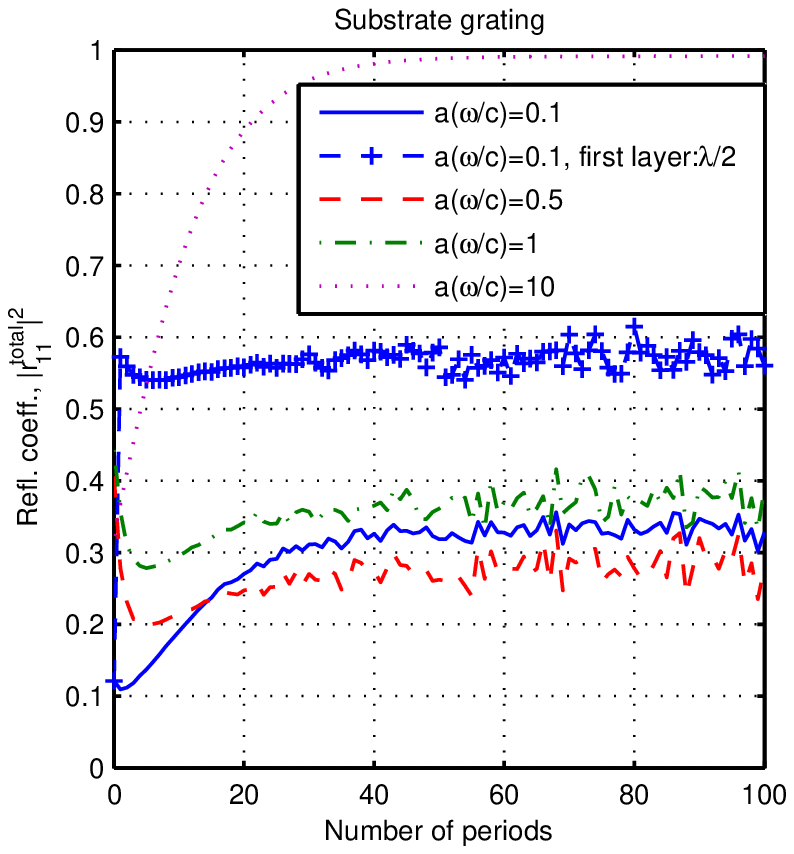}%
\caption{The reflectivity $\left|r_{11}^\text{total}\right|^2$ as a function of the number of periods in the \algaas{}/GaAs Bragg grating, for four waveguides of various width, $a$.}
\label{fig:reflection_substrate}
\end{figure}

We proceed to consider a Bragg grating in the substrate below the waveguide, as shown in Fig.~\ref{fig:transfer_geom} II). For nanowire applications, such structures can be realized by growing the substrate Bragg grating before the nanowire. This geometry has the advantage that it is easier to control the thickness and composition of the layers compared to the structure with the grating within the nanowire. The structure consists of the same materials as for the case with the grating within the waveguide, and we consider the same four waveguide widths as before. The half-width of the computational cell, $L$, used in the calculations was $1000/(\omega/c)$, which is larger than for the case with the grating within the waveguide. The reason for this is that the transition towards this substrate Bragg grating represents a more significant change in the modal fields. The coupling from the fundamental mode to higher order modes including radiation modes is therefore enhanced, and these modes will be more influenced by the artificial boundary conditions. The modal cut-off limit was taken to be $\beta_\text{lim}=3(\omega/c)i$. The resulting reflectivity as a function of the number of periods is shown in Fig.~\ref{fig:reflection_substrate}.

\begin{figure}%
\includegraphics[width=\columnwidth]{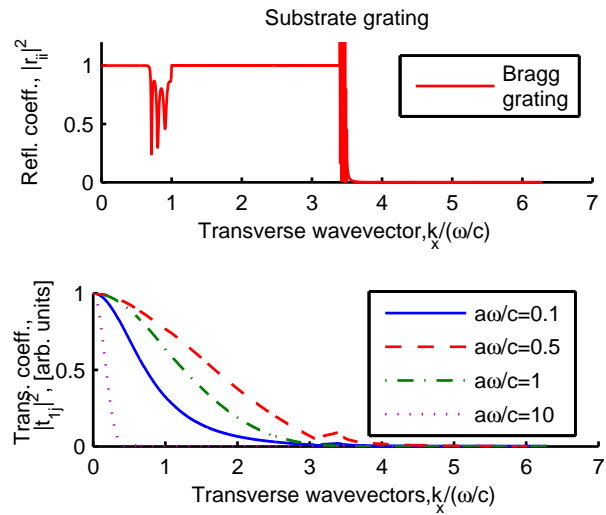}%
\caption{Transmission coefficients from the fundamental mode into the plane wave components in the first layer of the substrate grating (lower plot), and the respective reflectivity when these plane waves are reflected due to the grating (upper plot). Note that the reflectivity is higher than unity for spatial frequencies corresponding to propagating modes in GaAs, but evanescent modes in the \algaas{}. As no energy is transported by the evanescent modes, this does not violate energy conservation.}
\label{fig:grating_a_10_100_per}%
\end{figure}

In the geometric optics limit, i.e. as the normalized width of the waveguide increases, this substrate grating and a corresponding grating within the waveguide should approach each other. In this limit the reflection and transmission coefficients of the bound modes can be approximated by those of plane waves at a homogeneous interface \cite{svendsen:general}. Comparing Fig.~\ref{fig:reflection_substrate} and Fig.~\ref{fig:bragg_top_01}, we see that this approximation is indeed accurate for $a=10(\omega/c)$. However in the highly diffractive regime, there are large differences between the two Bragg geometries. Using a substrate grating, increasing diffraction will lead to decreased reflection. It is not possible to compensate for this by adding more layers. Fig.~\ref{fig:grating_a_10_100_per} helps us understand this effect. The upper plot displays the reflectivity of the quarter wave stack separately, i.e. the reflectivity of plane waves incident from the first substrate layer towards the remaining quarter wave stack. The lower plot displays the transmission coefficients from the fundamental mode of the waveguide into each of these plane waves.  The Bragg grating has reduced reflectivity in the region $k_x=0.75 (\omega/c)$ to $k_x=(\omega/c)$. Increasing the number of periods in the Bragg grating will increase the frequency of the oscillations in this region, but it will not decrease the width of this region with reduced reflectivity. The energy transmitted into plane waves in this low reflectivity region will therefore be partly transmitted through the grating and transported away. This explains why we do not achieve high reflectivity. For highly diffractive waveguides, a significant amount of the energy is transmitted into evanescent modes ($k_x>\omega/c$). The evanescent modes do not transport energy, so eventually they will couple back into propagating plane waves; especially to the plane waves with similar spatial frequencies. It is thus natural to assume that most of the energy in the evanescent modes is coupled back into the plane waves with reduced reflectivity; thus a large part is transported away from the structure. As $a\omega/c\to0$, the portion of the fundamental mode transmitted into the region with $k_x<0.75(\omega/c)$ will decrease, thus reducing the effect of the grating. 
 
For very small $a\omega/c$, the effective refractive index of the fundamental mode tends to the refractive index of the cladding medium (vacuum). The reflectivity of the fundamental mode will therefore converge to zero for very thin waveguides if the cladding material is the same as the substrate, as would be the case in the absence of a substrate grating. By adding a substrate layer of \algaas{} at the end of a GaAs waveguide, the effective index contrast for the fundamental mode will first decrease and then increase again as the waveguide width is decreased. As a consequence, the reflection coefficient for the fundamental mode towards the \algaas{} substrate will vary correspondingly. One may thus achieve relatively high reflectivity, but this is due to the fact that there is a high effective index contrast at the waveguide/substrate interface, not due to  constructive interference in the quarter wave grating.  

As a function of the number of periods, the reflectivity for the thinner waveguides decreases before it starts to increase (Fig.~\ref{fig:reflection_substrate}). This can be understood in terms of the phase shifts at the interfaces. As the effective refractive index of the waveguide decrease, the phase of the reflection at the interface between the waveguide and the substrate will change. As discussed in Sec.~\ref{sec:bragg_design}, this phase shift may lead to destructive interference between the backscattered contributions. For the two extreme waveguide widths $a=10(\omega/c)$ and $a=0.1(\omega/c)$, the phase shift of the first reflection coefficient is close to 0 or $\pi$, respectively. This leads to destructive interference for the waveguide width $a=0.1(\omega/c)$. To compensate for this, we doubled the thickness of the first layer in the Bragg grating, which strongly increased the reflectivity. The result is shown in Fig.~\ref{fig:reflection_substrate}.

\begin{figure}%
\includegraphics[width=\columnwidth]{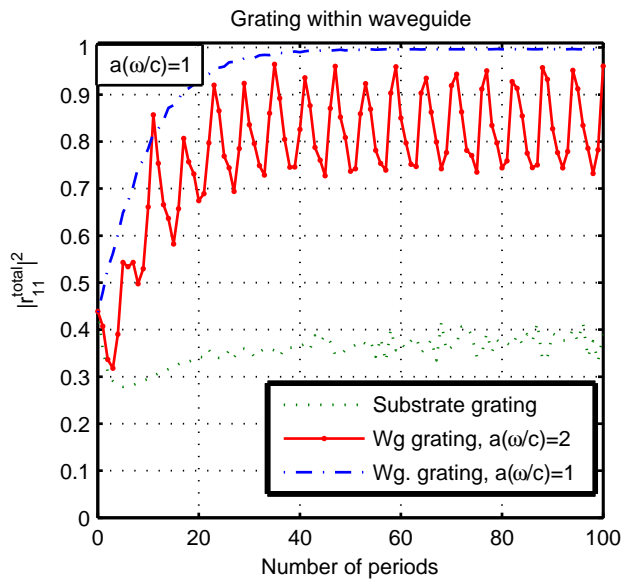}%
\caption{The reflection coefficient $\left|r_{11}^\text{total}\right|^2$ as a function of the number of periods in three GaAs/Al$_{0.3}$Ga$_{0.7}$As Bragg gratings. The central waveguide has width $a(\omega/c)=1$. The middle curve displays  the reflection properties of a grating of lateral width $a(\omega/c)=2$. The reflection coefficient for a substrate grating and a grating of the same width as the central waveguide is also shown for reference (taken from Fig.~\ref{fig:bragg_top_01} and Fig.~\ref{fig:reflection_substrate}).}
\label{fig:intermediate_a_1}%
\end{figure}

We have seen that the differences between the two grating geometries become large in the diffraction limit. Before we proceed to 2D calculations on nanowires, we therefore consider an intermediate geometry. Here, the segments of the grating have a larger lateral width than the central waveguide. Such structures can be realized by first growing a substrate Bragg grating, and then etch to reduce the lateral size of the reflector. This could be a potential way to overcome some of the weaknesses associated with substrate Bragg gratings, while maintaining a structure that is relatively easy to fabricate. The width of the central waveguide in the calculation was taken to be $a(\omega/c)=1$. Fig.~\ref{fig:intermediate_a_1} displays the reflection coefficient $\left|r_{11}^\text{total}\right|^2$ as a function of the number of periods for a grating of lateral width $a(\omega/c)=2$. The simulation parameters were $\beta_{\text{lim}}=5(\omega/c)i$, $\gamma_i=0.1$, and $L=100/(\omega/c)$. As a reference, we also show the corresponding reflection coefficient for a substrate grating and a grating with equal width as the central waveguide (taken from Fig.~\ref{fig:bragg_top_01} and Fig.~\ref{fig:reflection_substrate}). 

From Fig.~\ref{fig:intermediate_a_1}, we see that the reflection of the $a(\omega/c)=2$ grating is intermediate between the two geometries discussed earlier. There are also large fluctuations as a function of the number of periods. In the transition from the central waveguide ($a(\omega/c)=1$) towards the segment with $a(\omega/c)=2$, the field of the fundamental mode experiences a significant alteration. This leads to large coupling into several higher order modes of the $a(\omega/c)=2$ waveguide. The quarter wave resonance condition is however only fulfilled for the fundamental mode. The total reflection coefficient $\left|r_{11}^\text{total}\right|^2$ will therefore oscillate as a function of the number of periods, depending on the interference conditions for the energy transmitted into the high number of higher order modes. 

For grating structures of increasing lateral width, the wavevector separation between neighboring modes will decrease. The fluctuations will therefore be smoothed out in the limit of very wide grating structures. 

\section{Results, nanowire structure}
\label{sec:results_nanowire}
In this section we consider the effect of Bragg reflectors at the end facet of a semiconductor nanowire. A GaAs nanowire will be used as an example, with a Bragg grating consisting of GaAs/\algaas{}. The nanowires have a hexagonal cross-section. The lateral size of the waveguide is described using the effective radius $\rho_\text{eff}$, defined such that a hexagon with effective radius $\rho_\text{eff}$ has the same area as a circle with radius $\rho_\text{eff}$.

\begin{figure}%
\includegraphics[width=\columnwidth]{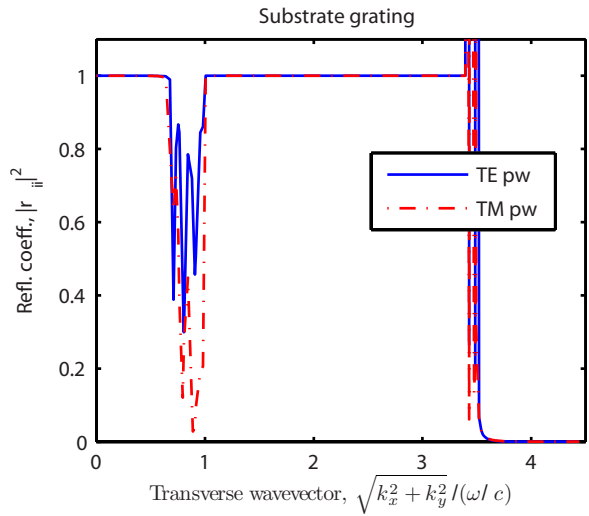}%
\caption{Reflection coefficients, $|r_{ii}|^2$, for plane wave components in an \algaas{} substrate. The plane waves are incident toward a 100 period GaAs/\algaas{} Bragg grating terminated by air.}%
\label{fig:2dBragg}%
\end{figure}

Substrate Bragg gratings were shown in the previous section to be inefficient for highly diffractive planar GaAs/AlGaAs waveguides. This was because only a small portion of the fundamental mode is transmitted into propagating plane waves with high reflectivity when diffraction effects are significant. In a 2D waveguide there will in general be coupling between modes of various polarization, thus both the TE and TM plane waves of the substrate have to be taken into consideration.  Fig.~\ref{fig:2dBragg} shows the reflection spectrum for TE and TM plane waves incident towards a substrate Bragg grating. For both polarizations there exists a region with reduced reflectivity for transverse wavevectors between $0.75(\omega/c)$ and $(\omega/c)$. This is a strong indication of the limited effect of Bragg reflectors in the substrate. A further study of Bragg gratings in the substrate is therefore omitted. Note however that due to the Brewster effect one might achieve very low reflectivity for TM polarized light in the absence of a Bragg grating. A Bragg grating may therefore help somewhat in those cases. 

\begin{figure}%
\includegraphics[width=\columnwidth]{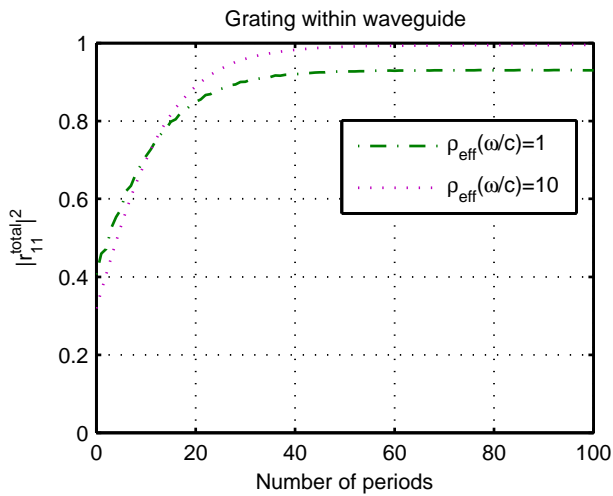}%
\caption{The reflection coefficient $\left|r_{11}^\text{total}\right|^2$ as a function of the number of periods in the GaAs/Al$_{0.3}$Ga$_{0.7}$As Bragg grating, for three waveguides of various width, $a$. }%
\label{fig:wg_wg_2_d}%
\end{figure}

Bragg gratings within the waveguide were seen to be promising in the planar case. The analysis is therefore extended to find the reflection properties of a hexagonal nanowire using such a grating. The two alternating materials are again taken to be \algaas{} and GaAs. The modes of the hexagonal waveguides were found using Comsol Multiphysics$^\text{TM}$. When extending from a 1D to a 2D analysis, there is a large increase in computational resources. We have therefore chosen to limit the 2D calculations to a more qualitative analysis, i.e. the simulation parameters are such that some inaccuracy should be expected in the results; we limit the number of modes included in the calculation and reduce the computational cell size. Two waveguide widths have been considered; $\rho_\text{eff}\omega/c=1$ and $\rho_\text{eff}\omega/c=10$. In the simulation we used $L=25$ and $\gamma=0.1$; $\beta_\text{lim}=(\omega/c)$ for $\rho_\text{eff}\omega/c=10$ and $\beta_\text{lim}$=0 for $\rho_\text{eff}\omega/c=1$.
Fig.~\ref{fig:wg_wg_2_d} displays the reflectivity of the fundamental mode as a function of the number of layers in the grating. 
The behavior is very similar to what was seen for the planar case. The grating does yield a significant increase in reflectivity. For the largest waveguide, the reflectivity of the fundamental mode as a function of the number of periods is in fact almost identical to what was seen for the corresponding planar waveguide. For the smaller waveguide, where diffraction effects are more pronounced, the reflectivity converges towards a value around 0.93. The deviation from the convergence limit of the corresponding planar structure is within the uncertainty due to the limited simulation parameters.

The maximum width for single mode operation in the (ZB) GaAs nanowire is around $\rho_\text{eff}(\omega/c)$=0.7. This implies an efficient radius of 97~nm when the excitation light is at the lasing wavelength $\lambda$=870 nm. The corresponding length of one period in the grating would be 237 nm, and a grating of 20 periods would thus be 4.7 $\muu${m} long. Such long nanowire gratings might be challenging to achieve with today's technology. For practical purposes one is thus limited to grow much shorter gratings. Higher reflectivity for shorter gratings may be achieved by increasing the index contrast, as is e.g. clearly seen in Fig.~\ref{fig:bragg_top_a1_large_contrast}. An increase in the aluminum composition up to $x=0.7$ would yield a refractive index of 3.15. In a planar structure with $a(\omega/c)=1$, such a grating would be capable of achieving a reflectivity of 0.9 after 7 periods, compared to 16 periods for $x=0.3$. A similar increase is to be expected for nanowire structures. It might be possible to also perform wet etching of the AlGaAs layers, as has been successfully done for VCSELs \cite{Choquette,Macdougal}. Wet etching of AlGaAs with high aluminum content will increase the refractive index contrast further, the refractive index of the oxidized AlGaAs layer is around 1.6 \cite{Kish}. A reflector consisting of such oxidized layers would however be non-conductive. Wet etching would thus be challenging for electrically driven nanowire lasers, as the end facets cannot be used for current injection. 
\section{Conclusion}
A semi-analytical model has been used to analyze the reflection properties of Bragg reflectors to increase the end facet reflectivity of diffractive waveguides. Such grating are promising to enable high quality nanowire laser cavities. We have considered a geometry with the grating within the waveguide/nanowire itself and a geometry with a substrate grating. The substrate grating has the advantage that the composition and thickness are more easily controlled, compared to the grating within the waveguide. For diffractive waveguides they were however found to yield a surprisingly small reflectivity. On the other hand, using the geometry with the grating within the waveguide, one could obtain near unity reflectivity even for extremely small waveguides, where only a small fraction of the field is within the waveguide. This would however imply using a high number of periods.

The semi-analytical model enables us to understand the mechanisms governing the efficiency of reflection gratings in connection to diffractive waveguides. The model is however not able to give very exact results for two dimensional highly diffractive waveguides, unless a high number of radiation modes and evanescent modes are included.

The structure with the grating within the waveguide is clearly seen to be the most promising when diffraction is significant. For GaAs waveguides terminated by such GaAs/\algaas{} gratings, one obtains maximum reflectivity after approximately 40 periods of the GaAs/\algaas{} grating, both for the planar waveguides and the nanowire waveguides. This is a high number that is not easily achieved with today's technology. To reduce the number of periods in the grating, one might increase the refractive index contrast. For the example considered here this could be achieved by increasing the aluminum composition in the AlGaAs layers, possibly in combination with wet etching. This will create a steeper increase in reflection as a function of the number of periods, but will also somewhat reduce the maximum obtainable reflectivity. 

\begin{acknowledgments}
This work was supported by the "NANOMAT" program (Grant No. 182091) of the Research Council of Norway. 
\end{acknowledgments} 
\appendix
\section{Reflection and transmission at boundary between waveguides}
\label{sec:bragg_within}

\begin{figure}%
\includegraphics[width=\columnwidth]{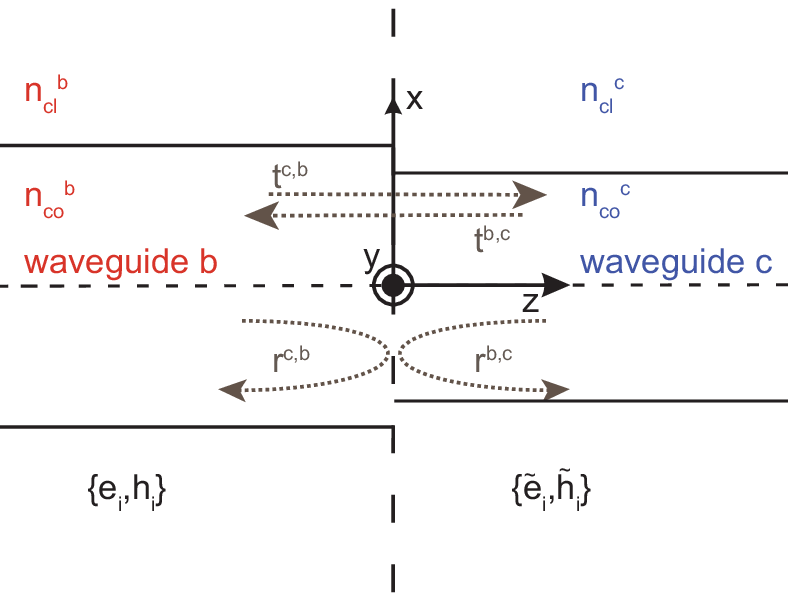}%
\caption{Geometry for reflection between two waveguides $b$ and $c$.}%
\label{fig:wg_wg_interface}%
\end{figure}

This section describes the procedure for calculating the reflection and transmission matrices at a boundary between two waveguides. The situation is sketched in Fig.~\ref{fig:wg_wg_interface}. Let waveguide $b$ be in the half-space $z<0$, and waveguide $c$ be in the half-space $z>0$. The electromagnetic field is discretized at both sides of the interface using the waveguide modes. Note that the modes can be divided into two parts; the discrete bound modes, and the continuous radiation modes. An artificial boundary condition e.g. periodic or metallic has to be applied to both half-spaces in order to fully discretize the modal spectrum. The modal field of waveguide $b$ is described by the set $\{\mathbf{e_i},\mathbf{h_i}\}$, where $\mathbf{e_i}=\mathbf{e_i}(x,y)$ and $\mathbf{h_i}=\mathbf{h_i}(x,y)$ are the electric and magnetic fields, respectively, of mode $i$. Similarly, the modal field of waveguide $c$ is described by the set $\{\mathbf{\tilde{e}_i},\mathbf{\tilde{h}_i}\}$.
The reflection and transmission matrices at the interface can now be found using the boundary conditions. Assume that mode $i$ of waveguide $b$ is incident from the left, the continuity of the transverse electric field yields
\begin{align}
&\mathbf e_i^{(t)}+\sum_j r_{ji}^{cb}\mathbf e_j^{(t)}=\sum_{q} t_{qi}^{cb}\mathbf{\tilde{e}}_{q}^{(t)},\label{wgwg_contE}%
\end{align}
valid for all $x$ and $y$. Here $r_{ji}^{cb}$ is the reflection coefficient from mode $i$ ($z<0$) to mode $j$ ($z<0$), when the mode is incident from waveguide $b$ toward waveguide $c$, and $t_{qi}^{cb}$ is the corresponding transmission coefficient from mode $i$ ($z<0$) to mode $q$ ($z>0$). The superscript $(t)$ stands for the transverse component ($x$ and $y$ components) of the vector. The boundary condition for the transversal magnetic field is similarly
\begin{align}
&\mathbf h_i^{(t)}-\sum_j r_{ji}^{cb}\mathbf h_j^{(t)}=\sum_{q} t_{qi}^{cb}\mathbf{\tilde{h}}_{q}^{(t)}\label{wgwg_contH}.
\end{align}
Eqs. \eqref{wgwg_contE} and \eqref{wgwg_contH} can be combined as follows. Take the vector product between \eqref{wgwg_contE} and $\mathbf{\mathcal{H}}_{m}^{(t)*}(x,y)$, and integrate over the unit cell. Similarly, take the vector product between $\mathbf{\mathcal{E}}_{m}^{(t)*}(x,y)$ and \eqref{wgwg_contH}, and integrate over the unit cell. Here, $\{\mathcal{H}_{m},\mathcal{E}_{m}\}$ belong to mode $m$ of an ambient dummy medium with refractive index $n_a$, according to Eq.~\eqref{eq:constvectors}. This yields
\begin{subequations}
\begin{align}
&\sum_j \left(\delta_{ji}+r_{ji}^{cb}\right)\Psi_{mj}=\sum_{q} t_{qi}^{cb}\tilde{\Psi}_{mq}\label{eq:E_bc}\\
&\sum_j \left(\delta_{ji}-r_{ji}^{cb}\right)\Phi_{mj}=\sum_{q} t_{qi}^{cb}\tilde{\Phi}_{mq}.\label{eq:H_bc}
\end{align}
\end{subequations}
Here, $\Psi_{mi}$, $\Phi_{mi}$ and $\tilde{\Psi}_{mi}$, $\tilde{\Phi}_{mi}$ are inner products, as given by Eq.~\eqref{eq:innerprod}, between the free space modes \eqref{eq:constvectors} and the modes of the waveguides, i.e. $\{\mathbf{e_i},\mathbf{h_i}\}$ and $\{\mathbf{\tilde{e}_i},\mathbf{\tilde{h}_i}\}$, respectively.
Let $t^{ab}=[t_{mi}]^{ab}$ and $r^{ab}=[r_{ji}]^{ab}$ be the shorthand notation for the reflection and transmission matrices for light incident from waveguide $b$ (with modal fields $\{\mathbf{e_i},\mathbf{h_i}\}$) towards the homogeneous ambient, and similarly for waveguide $c$ (with modal fields $\{\mathbf{\tilde{e}_i},\mathbf{\tilde{h}_i}\}$). From \eqref{eq:r_t_front} we have:
\begin{subequations}
\begin{align}
2t^{ab}&=\left([\Psi_{mi}]+[\Phi_{mi}]\right)-\left([\Phi_{mi}]-[\Psi_{mi}]\right)r^{ab}\\
2t^{ac}&=\left([\tilde{\Psi}_{mi}]+[\tilde{\Phi}_{mi}]\right)-\left([\tilde{\Phi}_{mi}]-[\tilde{\Psi}_{mi}]\right)r^{ac}\\
&\left([\Psi_{mi}]+[\Phi_{mi}]\right)r^{ab}=\left([\Phi_{mi}]-[\Psi_{mi}]\right)\\
&\left([\tilde{\Psi}_{mi}]+[\tilde{\Phi}_{mi}]\right)r^{ac}=\left([\tilde{\Phi}_{mi}]-[\tilde{\Psi}_{mi}]\right).
\end{align}
\end{subequations}
We eliminate $\left([\Psi_{mi}]-[\Phi_{mi}]\right)$ and  $\left([\tilde{\Psi}_{mi}]-[\tilde{\Phi}_{mi}]\right)$;
\begin{subequations}
\begin{align}
2t^{ab}\left(I-r^{ab}r^{ab}\right)^+&=\left([\Phi_{mi}]+[\Psi_{mi}]\right)\\
2t^{ac}\left(I-r^{ac}r^{ac}\right)^+&=\left([\tilde{\Phi}_{mi}]+[\tilde{\Psi}_{mi}]\right).
\end{align}
\end{subequations}
Eqs.~\eqref{eq:E_bc} and \eqref{eq:H_bc} can now be rewritten
\begin{subequations}
\begin{align}
&\left([\Phi_{mi}]+[\Psi_{mi}]\right)\left(I-r^{ab}r^{cb}\right)=\left([\tilde{\Phi}_{mi}]+[\tilde{\Psi}_{mi}]\right)t^{cb}\label{eq:s}\\
&\left([\Phi_{mi}]+[\Psi_{mi}]\right)\left(r^{ab}-r^{cb}\right)=\left([\tilde{\Phi}_{mi}]+[\tilde{\Psi}_{mi}]\right)r^{ac}t^{cb}\label{eq:ss}
.\end{align}
\end{subequations}
We solve \eqref{eq:s} and \eqref{eq:ss} for $r^{cb}$ and $t^{cb}$;
\begin{subequations}
\label{eq:wgc_wgb}
\begin{align}
&t^{cb}=G^{cb}\left(I-{r^{ab}}r^{cb}\right)\\
&r^{cb}=\left(G^{cb}-r^{ac}G^{cb}r^{ab}\right)^+\left(G^{cb}r^{ab}-r^{ac}G^{cb}\right)
.\end{align}
\end{subequations}
Here we introduced 
\begin{subequations}
\begin{align}
G^{cb}&\equiv\left([\tilde{\Phi}_{mi}]+[\tilde{\Psi}_{mi}]\right)^+\left([\Phi_{mi}]+[\Psi_{mi}]\right)\\
&=\left(I-r^{ac}r^{ac}\right)\left(t^{ac}\right)^+t^{ab}\left(I-r^{ab}r^{ab}\right)^+.\nonumber\\
\label{eq:G_bc}
\end{align}
\end{subequations}

The opposite reflection and transmission, $r^{bc}$ and $t^{bc}$, can be found in the exact same way. The result is immediately available by interchanging the indices $b$ and $c$, i.e.
\begin{subequations}
\label{eq:wgb_wgc}
\begin{align}
&t^{bc}=G^{bc}\left(I-r^{ac}{r^{bc}}\right)\\
&r^{bc}=\left(G^{bc}-r^{ab}G^{bc}r^{ac}\right)^+\left(G^{bc}r^{ac}-r^{ab}G^{bc}\right)
,\end{align}
\end{subequations}
where
\begin{subequations}
\begin{align}
G^{bc}=\left(G^{cb}\right)^+.
\label{eq:G_cb}
\end{align}
\end{subequations}
Using Eqs.~\eqref{eq:wgb_wgc} and Eqs.~\eqref{eq:wgc_wgb}, in combination with the recursive relations \eqref{eq:recursive}, the scattering properties of e.g. a complete Bragg structure can be calculated.

%

\end{document}